\documentclass[fleqn,11pt]{article}
\usepackage{psfig}
\title{Optical Pumping : Experiment and Theory Revisited}
\author{Zotin K.-H. Chu} 
\date{  
Physics Branch, Room W, 4/F, 16, Lane 21, \\ Guanghui Road , Wenshan District , Taipei, Taiwan 11646, China 
}
\begin{document}
\maketitle
\begin{abstract}
The objective of this paper is to share our enthusiasm for optical
pumping experiments  and to encourage their use in researches on
practical physics. The experimental technique has been well
developed and the apparatus sophisticated, but, by paying
attention to a few details, reliable operation can be repeated.
Some theoretical principles for optical pumping are also
introduced and they can be demonstrated experimentally.
\end{abstract}
\doublerulesep=7.2mm    %
\baselineskip=7.2mm
%
\bibliographystyle{plain}
\section{Introduction}
There are many methods developed to investigate the hyperfine
structure of alkali atoms, e.g., the classical and well-known
optical spectroscopy. However, it has the disadvantage that it is
necessary to evaluate a small quantity (hyperfine splitting or
hfs) as a difference of two large quantities (the optical
frequencies) [1]. Some techniques have been applied to study the
hfs with direct transitions between the hyperfine levels, i.e.,
magnetic-dipole radio-frequency transitions. As no such transition
can be detected, unless in the sample of atoms under investigation
the two hyperfine levels involved in the transition have
appreciably different occupation numbers, and since this does not
occur under thermodynamic equilibrium conditions (the hfs is $\ll
k T$, $k$ is the Boltzmann constant, $T$ is the absolute
temperature), other methods have been devised to alter these
occupation numbers. One of the most important and simple among
these techniques is the optical pumping [1-2]. Kastler was awarded
the Nobel Prize in 1966 since,  between 1949 and 1951, he was
supplementing the method by the the technique of optical pumping,
which makes it possible to apply 'optical methods for studying the
microwave resonances' to the fundamental states of atoms [3].
\newline
It was firstly proposed by Kastler in 1950 and has since then been
developed by many researchers and has many applications [3]. For
instance, Kastler and Brossel's groups discovered  numerous
phenomena related to high-order perturbations: multiple quantum
transitions, effects of Hertzian coherence, demonstration of
Hertzian resonance shifts under the influence of optical
irradiation, and profound modification of the properties of an
atom by the presence of a radio-frequency field [3] (before these
achievements, Brossel and Kastler together then proposed the {\it
double resonance method}, which combines optical resonance with
magnetic resonance). At the same time, with the same techniques,
other teams were achieving important results: measurement of
nuclear quadrupole electric moments of alkali metal atoms,
discovery of exchange collisions, displacement of hyperfine
resonances by collisions with molecules of a foreign diamagnetic
gas, and others. \newline In fact, Kastler showed that the optical
excitation of atoms with circularly polarized light made it
possible to transfer the angular momentum carried by the light to
the atoms and thus to concentrate them in the ground state, either
in the positive $m$ sublevels or in the negative $m$ sublevels
(depending upon whether the light is $s +$ or $s -$) and that it
was possible, by the optical pumping, to create an atomic
orientation and also, due to the coupling between the electronic
magnetic moment and the nuclear spin, a nuclear orientation. In
this manner, it should have been possible to obtain distributions
very different from the Boltzmann distribution and thus to create
conditions permitting the study of the return to equilibrium,
either by relaxation or under the influence of a resonant field
[3]. Here we shall devote some attention to the most important
aspects of the optical pumping. The experimental aspects will be
introduced firstly and then the simplified theoretical background
is demonstrated.\newline
\section{Optical Pumping Technique}
The technologies of optical pumping [1-3] and spin-exchange
optical pumping [4] have in recent years pervaded diverse areas of
research.
In the attempt to produce ever-larger quantities of noble gas with
high ($\ge 0.5$) polarizations, increasingly intense sources of
pump radiation have been used, evolving from the milliwatt alkali
discharge lamps of the earliest experiments [5]
to the $100$-W (or more) diode laser arrays used in much current
work [6]. \newline Note that the ultimate degree of polarization
that can be attained in an optical pumping experiment depends
critically on the relaxation rates in the ground state [1,3].
Pumping rates with conventional lamps rarely exceed a few thousand
photon absorptions per second; and, consequently, relaxation times
from collisions and other dissipative mechanism can be no shorter
than a few milliseconds if large degrees of polarization are to be
attained. The simplest is to place the atoms in a very large
evacuated container or atomic beam so that the time between
interatomic collisions or wall collisions is long.
\newline
Fig. 1 shows the schematic experimental arrangement. The lamp
creates the difference between the occupation numbers of the
hyperfine sublevels. The photomultiplier (PM) tube detects a
signal when the frequency of the applied radio frequency field is
in resonance with the energy difference between the two hyperfine
sublevels, and a standard servo system can lock the radio
frequency to the atomic transition. In order to have a light
source having different intensities on the hyperfine components,
one can use a filter or a very narrow tunable source (laser), but
this is not strictly necessary since ordinary spectral lamps
always have a nonuniform spectral distribution over a hyperfine
multiplet.
\newline
An example of a hyperfine multiplet is demonstrated in Fig. 2. If
the atoms are submitted to resonance radiation whose intensity is
frequently independent over the spectral region of absorption
({\it white} radiation) the ratio of the probability of absorption
to that of spontaneous emission is exactly the same for all
hyperfine components of the multiplet, and thus the ground-state
sublevels (initially equally populated) continue to be equally
populated. But if the incoming radiation is not white, and its
intensity on the hyperfine components connecting one hyperfine
state $|a\rangle$ of the ground state to the upper state is lower
than  the intensity on the other components connecting the other
state $|b\rangle$, the state $|b\rangle$ will be depleted at a
larger rate than it is refilled, whereas the opposite occurs to
the state $|a\rangle$.  One can thus alter the Boltzmann
distribution of the atoms in the ground state. If one destroys
this population difference with a radio-frequency transition
between the two hyperfine sublevels, the occupation number of the
less absorbing level $|a\rangle$ decreases, whereas that of the
more absorbing level $|b\rangle$ increases.
\newline
To summarize, optical pumping consists of depopulation pumping,
repopulation pumping, and relaxation [1]. Depopulation pumping
occurs when certain ground-state sublevels absorb light more
strongly than others. Since atoms are removed more rapidly from
the strongly absorbing sublevels, an excess population will tend
to build up in the weakly absorbing sublevels. Repopulation
pumping can occur when the atomic ground state is repopulated as a
result of spontaneous decay of a polarized excited state.
Relaxation can be caused by many mechanisms.  For example, the
most common one is collisions of the polarized atoms with other
atoms or molecules, collisions of polarized atoms with the
container walls, spatial diffusions of the polarized atoms from
regions of high polarization to region of lesser polarization, and
trapping of resonance radiation. In fact, optical pumping itself
can be viewed as a relaxation mechanism in which an ensemble of
atoms relaxes to a polarized steady state because of repeated
collisions with the polarized, directional, or frequency-selected
photons of pumping light. However, in most of optical pumping
experiments, the behavior of the pumped atoms is determined by
monitoring changes in the intensity of light that has interacted
with the atoms. There are, in general, two detection systems : (i)
fluorescence monitoring in which the fluorescent light emitted by
the atoms is observed, (ii) transmission monitoring in which a
probing external light beam is observed after the light beam has
passed through the vapor. The former provides direct information
about the polarization of he excited state while the latter gives
us direct information about the polarization of the ground state.
\section{Theoretical Aspects of Optical Pumping}
We firstly assume that the evolution of an individual atom (of the
vapor) is described by the Schr\"{o}dinger equation
\begin{equation}
 i \hbar \frac{\partial}{\partial t} |\psi\rangle = {\cal H} |\psi
 \rangle, \hspace*{12mm} {\cal H}\equiv{\cal H}_0 +V,
\end{equation}
where ${\cal H}_0$ is the same for all atoms (in the vapor) with
${\cal H}_0 |i\rangle = E_i |i\rangle$ ($|i\rangle$ being
eigenstates) and the small perturbation $V$ represents a randomly
fluctuating collisional interaction, an external radio-frequency
field, or other processes. Here, we are interested in a set of
ground-state basis functions (designated by $\mu$,$\nu$, etc.) and
a set of excited-state wave functions (designated by $m$,$n$,
etc.) considering the optical pumping measurements. In fact, we
define
\begin{equation}
 {\cal H}_g =\sum_{\mu,\nu} |\mu\rangle\langle \mu|({\cal
 H}-E_g)|\nu\rangle\langle\nu|,
\end{equation}
to be a ground-state Hamiltonian where $E_g$ is the mean energy of
the ground-state sublevels. The excited-state Hamiltonian ${\cal
H}_e$ can be defined in a similar way. Both ${\cal H}_g$ and
${\cal H}_e$ are traceless [1]. \newline Now, for an atom of
nuclear spin $I$ and electronic spin $J$, ${\cal H}_g$ and ${\cal
H}_e$ can be represented to sufficient accuracy by an effective
Hamiltonian of the form
\begin{displaymath}
 h\, A \,I \cdot J+h\, B \frac{[3(I\cdot J)^2+\frac{3}{2}(I\cdot
 J)-(I)(I+1)J (J+1)]}{2 I (2I-1)J(2J-1)}+
\end{displaymath}
\begin{equation}
 \hspace*{36mm} g_J \mu_0 J\cdot {\bf H}-\frac{\mu_I}{I} I\cdot {\bf H}.
\end{equation}
$A$ and $B$ are the magnetic dipole- and electric
quadrupole-interaction constants, respectively. $g_J$ and $\mu_I$
are the gyromagnetic ratio of the electronic spin $J$ and the
nuclear moment, respectively. The atoms are subjected to an
external magnetic field ${\bf H}$ here. \newline During the
optical pumping measurement, the signals observed usually are
proportional to the mean value $\langle M \rangle$ of some atomic
observable $M$ (say, certain component of the atomic angular
momentum). Normally we need to know the density matrix $\rho$ of
the atoms to calculate the average value of atomic observables.
$\rho=\sum_{i=1}^N |\psi_i\rangle\langle\psi_i|$ and the
probability of finding a given atom (of the vapor) in the sublevel
$|n\rangle$ is $\langle n |\rho|n \rangle$. Here, $|\psi_i\rangle$
is the wave function describing each atom (of the vapor). Thus,
with this, we have
\begin{equation}
 \langle M \rangle = \sum_n \langle n |\rho M|n
 \rangle=\frac{1}{N}\sum_{i=1}^N \langle \psi_i | M |
 \psi_i\rangle \equiv \mbox{Tr} [\rho M].
\end{equation}
Consequently, following the same procedures for optical pumping
experiments, we have
\begin{equation}
 \rho=\rho_g +\rho_e,
\end{equation}
with $\rho_g$ for the ground-state part and $\rho_e$ for the
excited-state part. \newline Next, we present the rate of change
of the density matrix which can be described by
\begin{equation}
 \frac{\partial}{\partial t} \rho=\frac{1}{i\hbar}[{\cal
 H}_0,\rho]+{\cal L} (\rho),
\end{equation}
which is known as the Liouville equation. The commutator $[{\cal
H}_0,\rho]$ follows directly from the equation (1). ${\cal
L}(\rho)$ denotes relaxation, pumping mechanisms, and all other
processes which ${\cal H}_0$ cannot prescribe. The details of
${\cal L}(\rho)$ could be traced in [1]. \newline In the following
step, we shall introduce the interaction representation to analyze
optical pumping measurements. The interaction-picture density
matrix $\sigma$ is defined by
\begin{equation}
 \sigma =\hat{\rho}=\exp(\frac{i {\cal H}_0 t}{\hbar})\, \rho \exp(\frac{-i {\cal H}_0
 t}{\hbar}).
 \end{equation}
Now, we then have
\begin{equation}
\frac{d}{dt} \sigma =\hat{{\cal L}} (\rho)
\end{equation}
which means once the interaction-picture density matrix is
constant there will be no pumping or relaxation mechanisms. Note
that, in terms of $\sigma$, we can also evaluate the average value
of any operator $M$ (cf. equation (4)). \newline Meanwhile we
frequently have [1]
\begin{equation}
 \frac{d}{dt} \sigma_{ij}=A_{ij}+\sum_k
 B_{ij}(\omega_k)\exp(i\omega_k \,t).
\end{equation}
It means the rate of change of the density matrix is the sum of
slowly varying part $A_{ij}$ and rapidly oscillating part
$B_{ij}(\omega_k)\exp(i\omega_k\,t)$. $\omega_k$ (the oscillating
frequency) is presumed to be very large compared to the components
of $A$ and $B$ : $\omega_k \gg A_{ij}$,  $\omega_k \gg
B_{ij}(\omega_r)$. One also presumes $A$ and $B$ to be of
comparable orders of magnitude and then the atomic evolution can
be derived by simply ignoring the rapidly oscillating terms and
solving the simplified equation
\begin{equation}
 \frac{d}{dt} \sigma_{ij}= A_{ij}.
\end{equation}
Above procedure is also known as the secular approximation [1] and
is useful in discussing the relaxation of an atomic
ensemble.\newline Finally, let us firstly assume that there are
$G$ sublevels of the atomic ground state. If the atoms of a vapor
were distributed at random among the ground-state sublevels, the
probability of finding the atom in any given sublevel will be
$1/G$, and the density matrix would be
\begin{equation}
 \rho_g =\frac{1}{G} \sum_{\mu} |\mu \rangle \langle
 \mu|=\frac{1}{G}.
\end{equation}
$\rho_g$ denotes a completely unpolarized ensemble. Next, we
define the polarization $P$ of an atomic vapor as the difference
between the actual density matrix and the density matrix of an
unpolarized ensemble. Thus, the polarization of the ground-state
density matrix $\rho_g$ is
\begin{equation}
 P_g =\rho_g -\frac{1}{G} \mbox{Tr} \,\rho_g.
\end{equation}
The excited-state polarization ($P_e$) can be defined in similar
manner (note that Tr $P$=0 [1]). People usually call the diagonal
matrix elements off the polarization 'population excesses'. Thus,
$\langle \mu|P_0|\mu\rangle$ is  the excess population of the
ground-state sublevel $\mu$ with respect to a random population
distribution. Off-diagonal components of the polarization or of
the density matrix are called coherence between the levels $\mu$
and $\nu$.\newline The polarization of the ground state is
$P=-{\cal H}_g/G k T$ in the absence of optical pumping or other
polarizing mechanisms where $T$ is the absolute temperature and
$k$ is the Boltzmann constant. We presume that the energy
splittings of the ground state are  much smaller than $kT$, so
that the thermal polarization is always very small. By optical
pumping the ground state, one can then produce much larger
polarization than the thermal polarization.\newline After these
experimental as well as theoretical explanations, we shall
illustrate again a simple optical pumping idea in Fig. 3. An atom
with a $^2S_{1/2}$ ground state and a $^2P_{1/2}$ excited state is
illuminated by circularly polarized resonance radiation which
propagates along the direction of a small magnetic field ($H$).
Ground-state atoms in the $+1/2$ sublevel cannot absorb light
since they cannot accomodate the additional angular momentum of
the photon in the $^2P_{1/2}$ excited state. However, ground-state
atoms in the $-1/2$ sublevel can absorb a photon and jump tot he
$+1/2$ sublevel of the excited state. Atoms in the $+1/2$
excited-state sublevel decay very rapidly and fall back to either
the $-1/2$ ground-state sublevel or the $+1/2$ ground-state
sublevel. The atom is twicea slikely to fall to the $-1/2$
sublevel as to the $+1/2$ sublevel, but, nevertheless, in the
absence of any relaxation mechanisms, all atoms will finally be
{\it pumped} into the $+1/2$ sublevel [1].
\section*{Optical Pumping without Collisions}
We noticed that, in his 1917 paper Einstein showed
\cite{Einstein:Therm} that even in the absence of collisions the
velocity distribution of a molecular gas takes on a Maxwellian
distribution due to the momentum transfer that takes place in the
absorption and emission of blackbody radiation. The absorbed and
emitted photons optically pump the rotational and vibrational
transitions, resulting in thermal distributions over the available
states. The rotational temperature of the CN molecule in
interstellar space \cite{M:Observ}, for example, is the result of
optical pumping by the cosmic microwave background-radiation
\cite{CMB:OP}.
\newline
\section{Conclusion}
The conventional optical pumping techniques are comprehensibly
introduced and described in experimental as well as theoretical
aspects. Several illustrations are demonstrated and these will be
useful to  those researchers in relevant field.


\newpage
\psfig{file=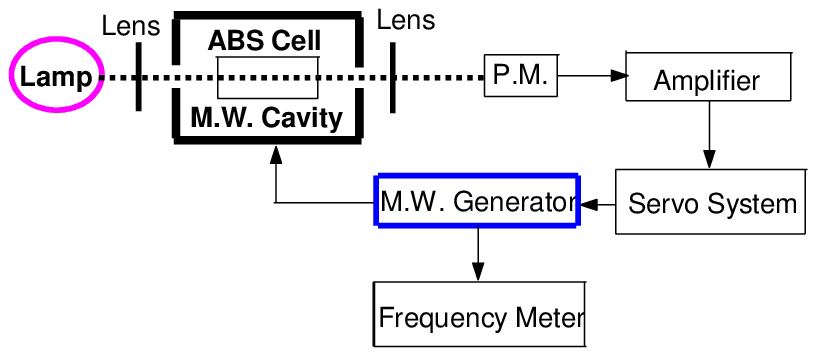,bbllx=2.5cm,bblly=13.0cm,bburx=14cm,bbury=28cm,rheight=16cm,rwidth=14cm,clip=}
%
\begin{figure}[h]
\hspace*{6mm} Fig. 1 \hspace*{1mm} Schematic experimental set-up
for investigating \newline \hspace*{7mm} hyperfine structures with
optical pumping approach [2].
\end{figure}

\newpage
\psfig{file=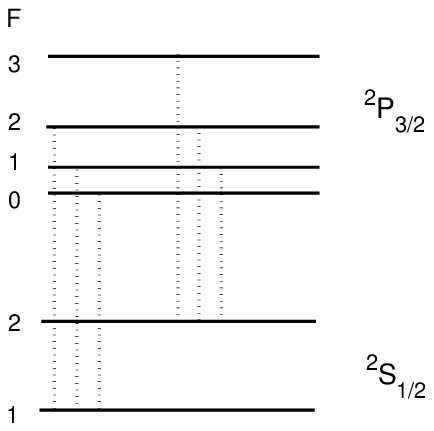,bbllx=0.1cm,bblly=18.0cm,bburx=14cm,bbury=28cm,rheight=10cm,rwidth=10cm,clip=}
%
\begin{figure}[h]
\hspace*{6mm} Fig. 2 \hspace*{1mm} Schematic hyperfine multiplet
corresponding to $D_2$ transition.
\end{figure}

\newpage

\psfig{file=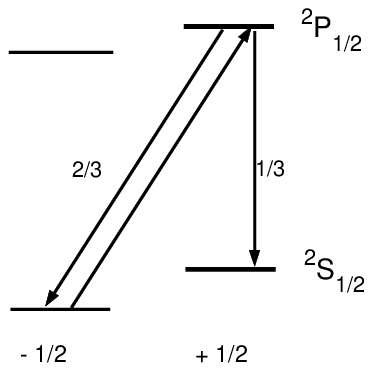,bbllx=0.1cm,bblly=18.0cm,bburx=14cm,bbury=28cm,rheight=10cm,rwidth=10cm,clip=}
%
\begin{figure}[h]
\hspace*{6mm} Fig. 3 \hspace*{1mm} Schematic idea of optical
pumping : depopulation pumping, \newline \hspace*{8mm}
repopulation pumping and relaxation [1]. All atoms will finally
\newline \hspace*{8mm} be {\it pumped} into the $+1/2$ sublevel [1].
\end{figure}
\end{document}